\documentclass[12pt]{article}
\usepackage{amsmath,cite}
\usepackage{amssymb}
\usepackage{slashed}
\usepackage[hidelinks]{hyperref}
\usepackage{cite}
\newcommand{\p}[1]{(\ref{#1})}
\newcommand{\be}{\begin{equation}}
\newcommand{\ee}{\end{equation}}
\newcommand{\bea}{\begin{eqnarray}}
\newcommand{\eea}{\end{eqnarray}}

\def\theequation{\arabic{section}.\arabic{equation}}
\begin{document}
\topmargin -1cm \oddsidemargin=0.25cm\evensidemargin=0.25cm
\setcounter{page}0
\renewcommand{\thefootnote}{\fnsymbol{footnote}} 
\begin{titlepage}
\vskip .7in
\begin{center}
{\Large \bf  Diagonalization Procedure for Reducible Massless ${\cal N}=1$ Supermultiplets  
 } \vskip .7in 
 {\Large 
Keith Glennon\footnote{e-mail: {\tt  keith.glennon@oist.jp }} and 
 Mirian Tsulaia\footnote{e-mail: {\tt  mirian.tsulaia@oist.jp}  }}
 \vskip .4in{  \it Okinawa Institute of Science and Technology, \\ 1919-1 Tancha, Onna-son, Okinawa 904-0495, Japan}\\
\vskip .8in
\begin{abstract}

We consider ${\cal N}=1$ supersymmetric systems
in $d=4$, $6$ and $10$ dimensions which consist
of reducible bosonic and fermionic massless representations of the Poincar\'e group.
 We show in detail how to decompose the corresponding Lagrangians into
 a sum of Lagrangians for irreducible representations of the 
 Poincar\'e group.
 We also outline a modification
of this procedure in the case of an anti-de Sitter background.

\end{abstract}

\end{center}

\vfill

\end{titlepage}

\renewcommand{\thefootnote}{\arabic{footnote}}
\setcounter{footnote}0

\newpage

\section{Introduction}

There are various approaches
to supersymmetric higher spin theories
both on flat and on anti-de Sitter ($AdS$) backgrounds (see \cite{Curtright:1979uz}--\cite{Kuzenko:1994dm}
for a pioneering work in this direction, and \cite{Vasiliev:1995dn}-- \cite{Sorokin:2017irs}
for reviews).
In most cases one considers irreducible representations
of the Poincar\'e or $AdS_d$ groups\footnote{ A possible implication of ${\cal N}=2$
supersymmetry in higher spin theories on $dS_4$ background
was recently discussed in \cite{Lang:2024dkt}.}, which in turn
form irreducible, and in some cases infinite-dimensional, 
representations of the corresponding supersymmetry algebra.
On the other hand, consideration of reducible representations might open some new interesting possibilities. Systems with reducible representations of the Poincar\'e group were studied  in detail in  
\cite{Francia:2002pt}--\cite{Campoleoni:2009gs}
and called either triplets or
generalized triplets, in the case fields with mixed symmetries are included.
 The main advantage of considering
reducible representations is that the corresponding free Lagrangians
and supersymmetry transformations have a relatively simple form
when compared to their irreducible counterparts.
This simplicity of the quadratic Lagrangians usually proves 
to be very helpful when constructing  cubic and higher order interactions, both for massive and massless higher spin fields
\cite{Metsaev:2007rn}--\cite{Metsaev:2020gmb}, \cite{Buchbinder:2021qrg}.

Following this line of thought, we elaborate on the ${\cal N}=1$ supersymmetric
model in $d=4,6,10$, space-time dimensions, constructed in \cite{Sorokin:2018djm}--\cite{Fotopoulos:2008ka}.
In this superstring-inspired model,
the fermionic sector consists of the fields described
by totally symmetric spin--tensors,
whereas
the bosonic sector consists of both totally symmetric fields
and fields with mixed symmetry of the simplest type\footnote{Supersymmetry in $d$ dimensions in the unfolded formulation was recently considered in \cite{Vasiliev:2025hfh}.}
$Y(n-1,1)$.
Because of the gauge invariant description,
the system contains, along with physical fields, also fields that are gauge degrees of freedom both in bosonic and in fermionic sectors.
One can show that, after the complete fixing of the gauge, the spectrum contains only massless reducible representations
of the Poincar\'e group 
described by
the Young diagrams $Y(n,0)$ and $Y(n-1,1)$
in the bosonic sector and Young diagrams $Y(n-1,0)$ in the fermionic sector. Both bosonic and fermionic representations
are reducible, since the traces of the fields are non-zero on-shell.
Moreover, the
corresponding ${\cal N}=1$ supermultiplets are irreducible in
$d=10$, but reducible in $d=4$ and $d=6$ also from the point of view
of representations of supersymmetry algebra.
As mentioned above, although this approach contains more fields
compared to an approach based on irreducible representations
of the supersymmetry and Poincar\'e groups,
 it allows one to construct Lagrangians for
both higher and lower spin fields  in a relatively simple
form. For example, in this way one can obtain: a new description
of linearized ${\cal N}=1$ supergravity and a linear
supermultiplet in four dimensions; linearized ${\cal N}=1$ supergravity and a
$(1,0)$ tensor multiplet in six dimensions 
and $N=1$ linearized supergravity in ten dimensions (see also \cite{Bertrand:2022pyi}
for a formulation of linearized $d=6$ supergravity in terms of reducible fields).

The purpose of the present paper is to develop a procedure
for the decomposition of fermionic and bosonic Lagrangians constructed in \cite{Sorokin:2018djm}
into a direct sum of Lagrangians for individual
irreducible representations of the Poincar\'e group \cite{Fronsdal:1978rb}--\cite{Fang:1978wz}.
This procedure will henceforth be referred to as a diagonalization procedure.
For totally symmetric representations
of the Poincar\'e group the diagonalization  has been carried out in
\cite{Fotopoulos:2009iw}-- \cite{Campoleoni:2012th}. 
Here we show how to perform the diagonalization procedure for fermionic totally symmetric fields and for the fields with   mixed symmetry\footnote{See \cite{Curtright:1980yk}-- \cite{Joung:2016naf} for different quadratic
 actions for mixed symmetry fields on Minkowski background.} of the type $Y(n,1)$. Let us note,
 that if  ${\cal N}=1$ supersymmetry is not required, both bosonic and fermionic Lagrangians, as well as the corresponding diagonalization procedures, are valid for any ($d \geq 4$) number of space-time dimensions.

The motivation behind developing the diagonalization
procedure is two-fold. Firstly,   although it is
more convenient to construct cubic and higher-order interaction vertices for reducible representations, in certain cases it might still be preferable to consider the interactions between reducible modes. The second reason 
is the possibility of addressing the question of a modification of the system constructed \cite{Sorokin:2018djm} to an $AdS_d$ background\footnote{The problem of a formulation of higher spin theories on backgrounds which are different from flat and $(A)dS_d$ spaces has been recently addressed in \cite{Tomasiello:2024jyu}.}.
Both deformation of bosonic and fermionic triplets that contain only totally symmetric fields to $AdS_d$
\cite{Sagnotti:2003qa}, \cite{Sorokin:2008tf}--\cite{Agugliaro:2016ngl}, 
as well as the relevant diagonalization procedure \cite{Fotopoulos:2010nj},
are known and can be applied directly to the present analysis.
However, the presence of
mixed-symmetry fields in the spectrum
causes some difficulties, and we shall comment on these
at the end of Section \ref{secdiag2}.

The paper is organized as follows.
In Section \ref{secreducible} we summarize the main
results of \cite{Sorokin:2018djm} which are needed
for the purpose of this paper.

In Section \ref{secdiag1} we demonstrate in detail the  diagonalization procedure for the simplest case of $n=1$, where the physical 
fields in the bosonic sector are  second-rank symmetric 
and antisymmetric tensors, whereas in the fermionic 
sector one has  first-rank spinor-tensor.
We also discuss various ways to deform  this system to an  $AdS_d$ background.

In Section \ref{secdiag2} we repeat this procedure in the more complicated case of $n=2$.
We show that after  diagonalization of the bosonic
Lagrangian one obtains: the Fronsdal Lagrangian for the spin
$3$ field, i.e., for the irreducible representation
$Y(3,0)$; a Lagrangian for the mixed symmetry irreducible representation of type
$Y(2,1)$; 
and two Lagrangians for vector fields, which correspond to the traces
of abovementioned (initially reducible) representations.
In the fermionic sector
one obtains the fields with spins $\frac{5}{2}$, 
$\frac{3}{2}$ and $\frac{1}{2}$.

 The detailed study of these examples suggests a clear pattern of how to generalize the procedure 
  and allows to conjecture the general 
 form of the decomposition of the reducible fields
 that diagonalizes the relevant Lagrangians for an arbitrary
 value of $n$. These topics are discussed in Section \ref{sgen}.

We conclude with a discussion of some possible
applications of our results.

Finally, the Lagrangians for massless irreducible
bosonic and fermionic higher spin fields on $AdS_d$, and the corresponding algebra of the operators are collected in the Appendix.

\section{Reducible \texorpdfstring{${\cal N}=1$}{N=1} supermultiplets}
\label{secreducible}
\setcounter{equation}0
Let us briefly summarize the main features
of the ${\cal N}=1$  supersymmetric system 
 constructed in \cite{Sorokin:2018djm}.
To this end, it is convenient 
to  introduce
an auxiliary Fock space, spanned by
one set of commuting and one set of anticommuting  auxiliary creation and annihilation operators
\begin{equation} \label{osc}
[\alpha_\mu, \alpha^+_\nu] = \{\psi_\mu, \psi^+_\nu  \} =
\eta_{\mu \nu}, \quad \alpha_{\mu} | 0 \rangle=
\psi_{\mu} | 0 \rangle=0
\end{equation}
Let us consider the bosonic fields first, where  one has a physical field
$|\phi \rangle$ and a set of auxiliary fields. The fields
are of the type $| X^{(p,q)} \rangle$, where $p$ is the number of
the commuting creation operators,  and $q$, which can be either $1$ or $0$, is the number of anticommuting  creation operators.
Obviously, the fields are symmetrical with respect to
$p$-indices and no symmetry between $p$-indices and the $q$-index is implied.

In the bosonic sector one has
the vectors
\begin{equation} \label{f-1}
|\phi^{(n,1)} \rangle, \quad |D^{(n-2,1)} \rangle, 
\quad |A^{(n-1,0)} \rangle, \quad |B^{(n-1,0)} \rangle
\end{equation}
\begin{equation} \label{f-2}
|C^{(n-1,1)} \rangle, \quad |E^{(n,0)} \rangle,
\quad  |F^{(n-2,0)} \rangle
\end{equation}
The corresponding quadratic Lagrangian has the form
\begin{eqnarray} \label{NSNS}
-{\cal L}_{bos.}&=& \langle \phi | l_0 | \phi \rangle -  \langle D |l_0 | D\rangle + i  \langle B| l_0 | A\rangle - i  \langle A | l_0 | B\rangle \\ \nonumber
&+&  \langle C| l^+ | D\rangle + i  \langle E| l^+ | B \rangle -  \langle C| l | \phi \rangle 
- i \langle F | l| B \rangle \\ \nonumber
&+& i  \langle C | g^+|A \rangle + i  \langle E| g | \phi \rangle-i  \langle F | g| D \rangle \\ \nonumber
&+&  \langle D | l| C\rangle - i  \langle B | l | E \rangle - \langle \phi | l^+| C\rangle + i  \langle B | l^+ | F \rangle 
\\ \nonumber 
&-& i  \langle A| g| C \rangle 
- i  \langle \phi| g^+| E \rangle + i  \langle D | g^+ | F \rangle \\ \nonumber
&+& \langle C| C \rangle + \langle E|  E \rangle - \langle F|  F\rangle 
 \end{eqnarray}
 with d`Alembertian, divergence and gradient operators being
 \begin{equation}
 l_0 = p\cdot p, \quad
 l^\pm= \alpha^ \pm \cdot p, \quad 
 g^\pm= \psi^ \pm \cdot p, \quad p= -i \partial_\mu 
 \end{equation}
 The system described above is a particular example
 of a generalized triplet \cite{Sagnotti:2003qa},
 which contains
 both  totally symmetrical 
 and mixed symmetry reducible 
  representations of 
  of the  Poincar\'e group.
 The field content  \p{f-1}
  and the explicit form of the Lagrangian  \p{NSNS}
  are dictated
 by the requirement of ${\cal N}=1$ supersymmetry, as we shall see at the end of the present Section.
 
The Lagrangian \p{NSNS} is invariant under the gauge transformations
\be \label{NSNSGT} 
\delta | \phi \rangle = l^+ | \lambda \rangle  + i g^+ | \rho \rangle ,
\ee
$$
\delta | D \rangle = l | \lambda \rangle  + i g^+ | \sigma\rangle ,  \quad
\delta | A \rangle =- l^+ | \sigma \rangle  + l | \rho \rangle - | \tau \rangle ,  \quad
\delta | B\rangle =- g| \lambda \rangle  -i | \tau \rangle ,\\
$$
$$
\delta | C \rangle = l_0| \lambda  \rangle  + i g^+ | \tau \rangle , \quad
\delta | E \rangle = l_0| \rho \rangle  - l^+ | \tau \rangle ,  \quad
\delta | F \rangle = l_0| \sigma \rangle  -  l | \tau \rangle 
$$
where the oscillator content for the parameters of gauge transformations 
 is
 \be\label{loparameters}
 |\lambda^{(n-1,1)} \rangle, \quad |\rho^{(n,0)} \rangle, \quad
 |\tau^{(n-1,0)} \rangle, \quad |\sigma^{(n-2,0)} \rangle
 \ee

 It can be shown (see \cite{Sorokin:2018djm} for the details),
 that one can use the gauge invariance  \p{NSNSGT}
 and the equations of motion in order
 to eliminate all on-shell  fields in \p{f-1}--\p{f-2}, except
 $|\phi \rangle$. This vector
  obeys the massless Klein-Gordon equation, is transverse with respect to both sets of its indices and 
 contains
 the totally symmetric representations of the Poincar\'e
 group of type $Y(n+1,0)$,  fields
 with mixed symmetry of type $Y(n,1)$, as well as their traces. Therefore, one obtains irreducible massless
 symmetric fields with spins $n+1, n-1,...,1$ or $0$
 and irreducible massless mixed symmetry fields
 of types $Y(r, 1)$ with $r=n, n-2,...,1$ or $0$ (i.e., a vector).
 
 As one can see from the Lagrangian
 \p{NSNS}, the fields \p{f-2} have no kinetic terms and   can be expressed  via their own equations of motion
 \bea \label{aux-exp}
 | C \rangle &=& - l^+ | D \rangle + l | \phi \rangle - i g^+ | A \rangle \\ \nonumber
 | E \rangle &=& - i l^+ | B \rangle  - i g | \phi \rangle \\ \nonumber
| F \rangle &=& - i l | B \rangle  - i g | D \rangle 
 \eea
 Thus the system can be formulated only in terms of the fields \p{f-1} with the corresponding Lagrangian
 \begin{eqnarray} \label{NSNS-1}
{\cal L}_{bos.}&=&- \langle \phi | l_0 - l^+ l - g^+ g| \phi \rangle +  \langle D |2l_0 + l^+ l - g^+ g| D\rangle +
\langle M |l_0 | M\rangle
\\ \nonumber
&-&  \langle \phi| l^+ l^+ | D\rangle  +   \langle \phi| l^+ g^+ | M \rangle
   -
\langle D | g^+ \, l|M \rangle
\\ \nonumber
&-& 
 \langle D| l \, l | \phi \rangle  +   \langle M| l \, g | \phi \rangle
  -\langle M | l^+ g |D \rangle
 \end{eqnarray}
 where $|M \rangle = |B \rangle - i |A \rangle$.

The states  in the fermionic sector contain only the bosonic oscillators, defined in \p{osc}.
The fields, as well as a Fock vacuum, carry the corresponding $d$ -dimensional spinorial index,
which will not be written explicitly. The field content is
\begin{equation} \label{f-3}
|\Psi^{(n,0)} \rangle, \quad |\chi^{(n-1,0)} \rangle, 
\quad |\Sigma^{(n-2,0)} \rangle
\end{equation}
with the corresponding Lagrangian
\begin{eqnarray} \label{L-F}
{ \cal L}_{f.}&=&-\langle \Psi |\gamma \cdot p |\Psi \rangle
+ \langle \Sigma |\gamma \cdot p |\Sigma \rangle
- \langle \Psi | l^+  |\chi \rangle
+ \langle \Sigma |l |\chi \rangle
  \\ \nonumber
&-& \langle \chi |l |\Psi \rangle
+ \langle \chi |l^+  |\Sigma \rangle
- \langle \chi |\gamma \cdot p |\chi \rangle,
\end{eqnarray}
where $\gamma^\mu$ are $d$-dimensional gamma matrices, with
$
\{ \gamma^\mu, \gamma^\nu \} = 2 \, \eta^{\mu \nu}
$.
 The Lagrangian
 \p{L-F}
 is invariant under gauge transformations
\be
\delta |\Psi \rangle = l^+ |
 \xi \rangle, \quad
\delta |\chi \rangle = -\gamma \cdot p 
| \xi \rangle  \quad
\delta |\Sigma \rangle = l 
| \xi \rangle 
\ee
where 
$| \xi^{(n-1,0)} \rangle$ is
a parameter.
Again, one can show,
that it is possible to gauge away the 
non-physical fields $|\chi \rangle$ and $|\Sigma \rangle$, and
one is left on-shell
 with the physical field  $|\Psi \rangle$. This field 
 obeys the massless Dirac equation, is transverse and
 describes a massless field with spin $n+ \frac{1}{2}$ as well as fields with lower spins $ n- \frac{1}{2},...,\frac{1}{2} $, (i.e., irreducible representations of the  Poincar\'e group)
  since no zero gamma--trace condition is implied.

Finally, one can establish ${\cal N}=1$ supersymmetry transformations
\begin{eqnarray} \label{susy-f}
\delta | \Psi \rangle &=& -(\gamma \cdot p) (\gamma \cdot \psi) \varepsilon | \phi \rangle + i \varepsilon | E \rangle
\\ \nonumber
\delta | \Sigma \rangle &=& -(\gamma \cdot p) (\gamma \cdot \psi) \varepsilon | D \rangle + i \varepsilon | F \rangle
\\ \nonumber
\delta | \chi \rangle &=&  (\gamma \cdot \psi) \varepsilon | C \rangle 
\end{eqnarray}
\bea \label{susy-b}
\delta \langle \phi |& = & \langle \Psi| (\gamma \cdot \psi)
\varepsilon \\ \nonumber
\delta \langle D |& = & \langle \Sigma| (\gamma \cdot \psi)
\varepsilon, \quad
\delta \langle B | =  \langle \chi|
\varepsilon \\ \nonumber
\delta \langle C |& = & \langle \chi| (\gamma \cdot p)(\gamma \cdot \psi) \varepsilon \\ \nonumber
\delta \langle A |& = & \delta \langle E |=
\delta \langle F |=0
\eea
which leave the sum of the Lagrangians \p{NSNS} and \p{L-F}
invariant.
One can check that the supersymmetry algebra
 close on-shell in the dimensions
$d=4,6$ and $10$.
Expressing the auxiliary fields via their own equations
of motion \p{aux-exp}, one can  can write the supersymmetry  transformation rules
for fermions as
\begin{eqnarray} \label{susy-f-1}
\delta | \Psi \rangle &=&- \left ( (\gamma \cdot p) (\gamma \cdot \psi) - g \right ) \varepsilon | \phi \rangle + l^+ \varepsilon | B \rangle
\\ \nonumber
\delta | \Sigma \rangle &=& 
- \left ( (\gamma \cdot p) (\gamma \cdot \psi) - g \right ) \varepsilon | D \rangle + l \varepsilon | B \rangle
\\ \nonumber
\delta | \chi \rangle &=&  l (\gamma \cdot \psi) \varepsilon | \phi \rangle -
l^+ (\gamma \cdot \psi) \varepsilon | D \rangle - i (\gamma \cdot p)
\varepsilon |A \rangle
\end{eqnarray}
As we mentioned earlier, the 
propagating  bosonic and fermionic degrees of freedom
are contained in  $|\phi \rangle$
and $|\Psi \rangle $ respectively, the rest of the fields in \p{f-1},  \p{f-2} and  \p{f-3}
are auxiliary.
For  $n=0$ the spectrum contains a  vector field
in the bosonic sector and a spin $\frac{1}{2}$
field in the fermionic sector,
i.e., one has an on-shell ${\cal N}=1$  Maxwell supermultiplet
in $d=4,6$ and $10$ dimensions.
The choice $n=1$ gives ${\cal N}=1$ on-shell supergravity and matter
supermultiplets, listed in the Introduction.
This pattern (i.e., supermultiplets with $n$ commuting and one anticommuting creation operator in the  bosonic field
and $n$ commuting creation operators in the fermionic field)
holds for an arbitrary value of $n$. 
The set of  auxiliary fields 
in \p{f-1},  \p{f-2} and  \p{f-3}
is further determined by the requirement of gauge invariance and
by ${\cal N}=1$ supersymmetry of the total Lagrangian.

Let us also note  that
the closure of the supersymmetry algebra, as well as
 the consistency of the equations
\p{aux-exp} with the transformation rules \p{susy-f}--\p{susy-b},
requires the use of fermionic field equations, derived from 
the Lagrangian \p{L-F}. This should have been expected since we are considering on--shell supersymmetry.

\section{Diagonalization for
\texorpdfstring{$n=1$}{n=1}} \label{secdiag1}
\setcounter{equation}0
\subsection{Flat space-time}
In order to demonstrate the details of the
diagonalization procedure and to see the pattern,
we start with the simplest case of $n=1$.

After performing the first step
of eliminating auxiliary fields in terms of their
own equations of motion, 
the bosonic Lagrangian \p{NSNS-1} reads
\be \label{n1b}
{\cal L}_{bos.} =   \phi^{\mu, \nu} \Box \phi_{\mu, \nu} -
M \Box M 
+2 \partial^\mu \phi_{\mu, \nu} \partial^\nu M
- \phi^{\tau, \nu} \partial_\tau \partial^\mu 
\phi_{\mu, \nu} - 
\phi^{\mu, \tau} \partial_\tau \partial^\nu 
\phi_{\mu, \nu}
\ee
The second step is to decompose the field
$\phi_{\mu, \nu}$ in terms of the fields with definite symmetries  
\be
\phi_{\mu, \nu} = \tilde \phi_{\mu \nu} + {\cal B}_{\mu \nu}, 
\quad \tilde \phi_{\nu \mu} = \tilde \phi_{\mu \nu}, \quad
{\cal B}_{\mu \nu}= - {\cal B}_{\nu \mu}
\ee
The Lagrangian \p{n1b}
written in terms of the new fields has the form
\be \label{sum}
{\cal L}_{bos.} = {\cal L}_{bos.}^\prime + {\cal L}_{bos.}^{\prime \prime}
\ee
where
\bea
&&{\cal L}_{bos.}^\prime=  \tilde \phi^{\mu \nu} \, \Box \, \phi_{\mu \nu}
- 2
\tilde \phi^{ \mu \tau } \partial_\tau \partial^\nu 
\tilde \phi_{ \mu \nu}
+ 4 \tilde D \, \partial^\mu \partial^\nu \,
\tilde \phi_{\mu \nu}
-
4 \tilde D \, \Box \, \tilde D 
\\ 
&&{\cal L}_{bos.}^{\prime \prime}=  {\cal B}^{\mu \nu} \, \Box \,  {\cal B}_{\mu \nu}
-2{\cal B}^{\mu \nu} \, \partial_\nu \partial^\tau \,  {\cal B}_{\mu \tau}  \label{C-R}
\eea
and we defined $\tilde D = - \frac{1}{2} M$.
 The final  step consists of splitting the
Lagrangian ${\cal L}_{bos.}^\prime$ into a sum of  Lagrangians
describing fields with  
spin $2$ and spin $0$. An obvious  candidate for 
the gauge invariant scalar is 
\be \label{db1}
{\Phi} = \tilde \phi^\mu{}_{\mu} - 2 \tilde D
\ee
For the spin $2$ field $\Phi_{\mu \nu}$ one takes an ansatz
\be \label{db2}
\tilde \phi_{ \mu \nu} = \Phi_{\mu \nu} + \frac{1}{d-2} \eta_{\mu \mu} \Phi
\ee
where the coefficient
in front of the second term 
is determined from the requirement
that the Lagrangian ${\cal L}_{bos.}^\prime$ splits into a sum of independent Lagrangians for the fields ${\Phi}_{\mu \nu}$ and ${\Phi}$.
Indeed, one can check  that this choice 
leads to the Lagrangian
\bea \nonumber \label{2+0}
{\cal L}_{bos.}^\prime &=&  \Phi^{\mu \nu} \, \Box \, \Phi_{\mu \nu} - 2\Phi^{\mu \tau} \partial_\tau \partial^\rho
\Phi_{\mu \rho}
-\Phi^{\mu}{}_{\mu} \, \Box \, \Phi^{\nu}{}_{\nu}+2
\Phi^{\rho}{}_{\rho} \, \partial_\mu \partial_\nu \, \Phi^{\mu \nu} \\ 
&+& \frac{1}{d-2} \Phi \, \Box \, \Phi
\eea
and this completes the treatment of the bosonic part of the system
for the case of $n=1$.

In the fermionic sector according to \p{L-F}
 one has the Lagrangian 
\be \label{lf-1}
{\cal L}_{f.} =  
 i \bar \Psi^\mu \gamma^\nu \partial_\nu \Psi_\mu
+ i \bar \Psi^\mu  \partial_\mu \chi
- i \bar \chi  \partial^\mu \Psi_\mu
- i \bar \chi \gamma^\nu \partial_\nu \chi
\ee
 invariant under the gauge transformations
\be
\delta \Psi_\mu = \partial_\mu \xi, \quad
\delta \chi = -\gamma^\mu \partial_\mu \xi
\ee
For the fermionic fields,
the first two steps of the diagonalization procedure 
 are not needed and 
one  starts from the third step.
Similarly to the case of the bosonic fields,  we introduce 
a spin $\frac{3}{2}$ field $\tilde \Psi_\mu$ and
a gauge invariant spin $\frac{1}{2}$ field 
$\tilde \Psi$
as
\be \label{df1}
\tilde \Psi_\mu =  \Psi_\mu + \frac{1}{2-d} \gamma_\mu \tilde \Psi,
\quad
\tilde \Psi = \gamma^\mu \Psi_\mu + \chi,
\ee
where the coefficient in front of the last term 
in the definition of the field $\tilde \Psi_\mu $
is chosen in such a way that the Lagrangian 
 \p{lf-1} written in terms
of the new fields is a sum of Fang -- Fronsdal Lagrangians
for spins $\frac{3}{2}$ and  $\frac{1}{2}$
\bea \label{lff-1} \nonumber
{\cal L}_{f.}&=& i \bar {\tilde \Psi}^\mu \gamma^\nu \partial_\nu \tilde \Psi_\mu 
-
i \bar {\tilde \Psi}^\mu  \partial_\mu 
\gamma^\nu \tilde \Psi_\nu 
- i \bar {\tilde \Psi}^\mu \gamma_\mu \partial_\nu \tilde \Psi^\nu 
+i (\bar {\tilde \Psi}^\mu \gamma_\mu) \gamma^\nu \partial_\nu (\gamma^\rho \tilde \Psi_\rho) \\ 
&+&\frac{i}{d-2} 
\bar {\tilde \Psi} \gamma^\mu \partial_\mu \tilde \Psi 
\eea
Finally, using the equations \p{db1}, \p{db2}, \p{df1}, as well as
\p{susy-b} and \p{susy-f-1}, it is straightforward to rewrite
the supersymmetry transformations in terms of irreducible modes.
From the field content given above, one can see that
this system describes:
 a linearized ${\cal N}=1$
supergravity and the linear multiplet in case of
$d=4$; a 
linearized ${\cal N}=1$ supergravity  and $(1,0)$ tensor multiplet in case of $d=6$;  and a linearized ${\cal N}=1$ supegravity  in 
case of $d=10$ dimensions \cite{Sorokin:2018djm}.

\subsection{Deformation to \texorpdfstring{$AdS$}{AdS}  space}
\label{adsdeformation-s}
In order to perform a deformation of the bosonic \p{NSNS} and
fermionic 
\p{L-F}
Lagrangians to an $AdS_d$ background, one can proceed in two different ways.
The first option is to start with the already decomposed
system described by a sum of the Lagrangians
\p{C-R}, \p{2+0} and \p{lff-1}. Then we perform its deformation
to $AdS_d$  space by replacing the usual partial derivatives with the 
covariant ones and adding the corresponding mass-like terms.
The  Lagrangians for  totally symmetric fermionic and bosonic fields can be directly read off from the equations \p{Fronsdal} and \p{Fang-Fronsdal}
which we give here for  convenience
\bea \label{2+0-ads}
{\cal L}_{bos.}^\prime &=&  \Phi^{\mu \nu} \, \Box \, \Phi_{\mu \nu} - 2 \Phi^{\mu \tau} \nabla_\tau \nabla^\rho
\Phi_{\mu \rho}
-\Phi^{\mu}{}_{\mu} \, \Box \, \Phi^{\nu}{}_{\nu}+2
\Phi^{\rho}{}_{\rho} \, \nabla_\mu \nabla_\nu \, \Phi^{\mu \nu} \\ 
\nonumber 
&+&\frac{1}{L^2} (2\, \Phi^{\mu \nu} \Phi_{\mu \nu} +(d-3)
\Phi^{\mu}{}_{\mu}\Phi^{\nu}{}_{\nu}) + \frac{1}{d-2}  (\Phi \, \Box \, \Phi
- \frac{1}{L^2} (6-2d) \Phi \Phi
 )
\eea
\bea \label{lff-1-ads} \nonumber
{\cal L}_{f.}&=& i \bar {\tilde \Psi}^\mu \gamma^\nu \nabla_\nu \tilde \Psi_\mu 
-
i \bar {\tilde \Psi}^\mu  \nabla_\mu 
\gamma^\nu \tilde \Psi_\nu 
- i \bar {\tilde \Psi}^\mu \gamma_\mu \nabla_\nu \tilde \Psi^\nu 
+i (\bar {\tilde \Psi}^\mu \gamma_\mu) \gamma^\nu \nabla_\nu (\gamma^\rho \tilde \Psi_\rho) \\ 
&+&\frac{i}{2L} ((d-2){\bar {\tilde \Psi}}^\mu {\tilde \Psi}_\mu -d \bar {\tilde \Psi}^\mu \gamma_\mu \gamma^\nu{ \tilde  \Psi}_\nu ) \\ \nonumber
&+&\frac{i}{d-2} 
(\bar {\tilde \Psi} \gamma^\mu \nabla_\mu \tilde \Psi +
\frac{(d-4)}{2L} \bar {\tilde \Psi}  \tilde \Psi)
\eea
The mass--like terms for
the spin $2$ and spin $\frac{3}{2}$ fields 
are determined from the requirement  
that  all gauge symmetries  of Lagrangians
\p{NSNS} and \p{L-F} are kept intact.
The mass--like terms for the gauge invariant scalar
and spin $\frac{1}{2}$ fields 
in \p{2+0-ads} and \p{lff-1-ads} are taken so that they
saturate the unitarity bounds for the
$AdS_d$ isometry group
$SO(d-1,2)$. These terms can also be  obtained
from diagonalization of the spin $2$ and spin
$\frac{3}{2}$ triplets on $AdS_d$.

For the antisymmetric field ${\cal B}_{\mu \nu}$
we get
\be \label{C-R-ads}
{\cal L}_{bos.}^{\prime \prime}=  {\cal B}^{\mu \nu} \, \Box \,  {\cal B}_{\mu \nu}
-2{\cal B}^{\mu \nu} \, \nabla_\nu \nabla^\tau \,  {\cal B}_{\mu \tau} +\frac{2 (d-2)}{L^2}
{\cal B}^{\mu \nu}   {\cal B}_{\mu \nu} = -\frac{1}{3} H^{\mu_1 \mu_2 \mu_3} H_{\mu_1 \mu_2 \mu_3}
\ee
where $H_{\mu_1 \mu_2 \mu_3} = \partial_{[\mu_1} {\cal B}_{\mu_2 \mu_3]}$.
In the case of $d=4$, one 
has another option for $AdS$ deformation.
Namely, one 
can first dualize the field
${\cal B}_{\mu \nu}$ to a scalar $a(x)$ as $\partial_{\mu} a \sim
\epsilon_{\mu \nu \rho \sigma} H^{\nu \rho \sigma}$
and then perform the $AdS_4$ deformation to a conformal scalar,
with the mass--like term being the same as in \p{2+0-ads}
for the scalar field $\Phi$. 
As a result,
one obtains  either the linear ${\cal N}=1$ supermultiplet  
\cite{Siegel:1979ai}--\cite{Ivanov:1980vb},  or
the Wess--Zumino supermultiplet, 
where the spin $\frac{1}{2}$ field being $\tilde \Psi$.
The requirement of ${\cal N}=1$ supersymmetry
for the case of four dimensions\footnote{
See \cite{Breitenlohner:1982bm}--\cite{deWit:2002vz} for a detailed discussion of various properties of the Wess--Zumino supermultiplet on $AdS_4$.}
fixes the  mass--like terms for the spin $0$ and the
spin $\frac{1}{2}$ fields as given in \p{2+0-ads} and \p{lff-1-ads}.
Finally, the fields  ${\Phi}_{\mu \nu}$ and $\tilde \Psi_\mu$, in turn form  a supermultiplet of the 
 ${\cal N}=1$ supergravity on $AdS_4$, constructed in \cite{Townsend:1977qa}.

As an alternative option one can perform the $AdS_d$
deformation of the Lagrangian \p{NSNS} from the beginning
\bea \label{t2}
{\cal L}_{bos.}&=&   \phi^{\mu, \nu} \Box \phi_{\mu,\nu } - B\Box A  - A \Box  B \\ \nonumber
& -&2E^\mu  \nabla_\mu B  - 2C^{\nu}\nabla^\mu \phi_{\mu,\nu } - 2C^\nu \nabla_\nu  A
- 2E^\mu\nabla^\nu
\phi_{ \mu, \nu } \\ \nonumber
&-&C^\nu C_\nu - E^\mu E_\mu\,
- \frac{1}{L^2} \Big (- (d-1)\phi^{\mu, \nu}  \phi_{\mu,\nu }
+ (d-3) \phi^{\mu, \nu}  \phi_{\nu,\mu} \\ \nonumber
&-&2 \phi^{\mu}{}_{,\mu} \phi^{\nu}{}_{,\nu}
-4 \phi^{\mu}{}_{,\mu} (A+B) - (3+d) (A+B)^2 \Big )
\eea
Using the commutation relations given in the Appendix
one can check that Lagrangian \p{t2} is invariant under the gauge transformations 
\bea \nonumber
&&\delta \phi_{\mu, \nu } = \nabla_\mu \lambda_\nu + \nabla_\nu \rho_\mu, \\ \nonumber
&&\delta A = - \nabla^\nu \rho_\nu - \tau, \\ \nonumber
&&\delta B= - \nabla^\nu \lambda_\nu+ \tau, \\
&&\delta C_\nu  =- \Box  \lambda_\nu 
+ \frac{1}{L^2}(d-1)
\rho_\nu
+ \nabla_\nu \tau, \\ \nonumber
&&\delta E_\mu=- \Box  \rho_\mu + \frac{1}{L^2}(d-1)
\lambda_\mu
-\nabla_\mu \tau.
\eea
Further steps are exactly the same as in the case of a flat background and lead to the Lagrangians \p{2+0-ads}
and \p{C-R-ads}. An analogous procedure, i.e., the diagonalization
of the fermionic triplets on $AdS_d$
to the spins $\frac{3}{2}$ and $\frac{1}{2}$ has been performed
in \cite{Agugliaro:2016ngl}, where further details  can be found.

\section{Diagonalization for 
\texorpdfstring{$n=2$}{n=2}} \label{secdiag2}
\setcounter{equation}0
The diagonalization procedure for the case $n=2$ essentially repeats the same steps as for $n=1$.
However, it has some novel features, such as  presence of the simplest mixed symmetry fields.

One starts from the Lagrangian \p{NSNS-1} which in this case reads
\bea \nonumber \label{ln2b-1}
{\cal L}_{bos.} &=&   
\phi^{\mu_1 \mu_2, \nu} \, \Box \, \phi_{\mu_1 \mu_2, \nu} -
\phi^{\mu_1 \mu_2, \nu_1} \partial_{\nu_1} \partial^{\nu_2}
\phi_{\mu_1 \mu_2, \nu_2}
- 2
\phi^{\mu_1 \mu_2, \nu} \partial_{\mu_2} \partial^{\mu_3}
\phi_{\mu_1 \mu_3, \nu} \\ 
&-&4 \phi^{\mu_1 \mu_2, \nu} \partial_\nu \partial_{\mu_1} M_{\mu_2}
+ 4 
\phi^{\mu_1 \mu_2, \nu} \partial_{\mu_1} \partial_{\mu_2} D_{\nu} 
\\ \nonumber
&-&4  D^\nu \, \Box \,  D_\nu
+2 D^{\nu_1} \partial_{\nu_1} \partial_{\nu_2} D^{\nu_2}
+ 4  D^\nu \partial_\nu \partial_\mu M^\mu  
-2 M^\mu \Box M_\mu 
\eea
Next, one decomposes the field $\phi_{\mu_1 \mu_2, \nu}$ into a sum
\be \label{sn2}
 \phi_{\mu_1 \mu_2, \nu} = \tilde  \phi_{\mu_1 \mu_2 \nu}
+ {\cal B}_{\mu_1 \mu_2, \nu}
\ee
where the field $\tilde  \phi_{\mu_1 \mu_2 \nu}(x)$ is totally symmetric
\be \label{symmetric}
\tilde  \phi_{\mu_1 \mu_2 \nu} = \frac{1}{3}
( \phi_{\mu_1 \mu_2, \nu}+  \phi_{\mu_1 \nu, \mu_2} +  \phi_{\mu_2 \nu, \mu_2})
 \ee
and the field ${\cal B}_{\mu_1 \mu_2, \nu}$ has the mixed symmetry
of the type $Y(2,1)$
\be \label{mixedsymmetric}
{\cal B}_{\mu_1 \mu_2, \nu}= \frac{1}{3}
( 2 \phi_{\mu_1 \mu_2, \nu}-  \phi_{\mu_1 \nu, \mu_2} -  \phi_{\mu_2 \nu, \mu_2}), \quad
{\cal B}_{\mu_1 \mu_2, \nu} = {\cal B}_{\mu_2 \mu_1, \nu}, \quad
{\cal B}_{(\mu_1 \mu_2, \nu)} =0
\ee
Introducing the linear combinations
\be
\tilde D_\mu = \frac{1}{3} (D_\mu - M_\mu), \quad
{\tilde M}_\mu = M_\mu + 2 D_\mu
\ee
and inserting them  together with \p{sn2} into
 the Lagrangian \p{ln2b-1}, one
 finds that it splits into two components
 \p{sum}.
The first component
${\cal L}_{bos.}^\prime$
describes simultaneously spin $3$ and spin $1$
fields and is given by the Lagrangian
\bea \label{tr-3-1}
{\cal L}_{bos.}^\prime &=&  \tilde \phi^{\mu_1 \mu_2 \mu_3} \, \Box \, \tilde \phi_{\mu_1 \mu_2 \mu_3} - 3
\tilde \phi^{\mu_1 \mu_2 \nu_1}
\partial_{\nu_1} \partial^{\nu_2}
\tilde \phi^{\mu_1 \mu_2 \nu_2} \\ \nonumber
&-& 12 \tilde \phi^{\mu_1 \mu_2 \mu_3} \partial_{\mu_1} \partial_{\mu_2} \tilde D_{\mu_3}
+ 12  \tilde D^\mu \, \Box \, \tilde D_\mu
+6  \tilde D^{\mu_1}  \partial_{\mu_1} \partial_{\mu_2}
\tilde D^{\mu_2} 
\eea
being invariant under the gauge transformations
with a symmetric parameter ${\tilde \Lambda}_{\mu \nu}$
\be
\delta  \tilde \phi_{\mu_1 \mu_2 \mu_3} = \partial_{(\mu_1}
{\tilde \Lambda}_{\mu_2 \mu_3) }
, \quad 
\delta  \tilde D_{\mu} = \partial^\nu \tilde \Lambda_{\mu \nu},
\ee
The second term 
in \p{sum} describes 
a reducible representation of the Poincar\'e group
with the mixed symmetry
of the type $Y(2,1)$ 
\bea \nonumber \label{bpp}
{\cal L}_{bos.}^{\prime \prime} &=&  {\cal B}^{\mu_1 \mu_2, \nu} \, \Box \,  {\cal B}_{\mu_1 \mu_2, \nu} - 
{\cal B}^{\mu_1 \mu_2, \nu_1}
\partial_{\nu_1} \partial^{\nu_2}
{\cal B}_{\mu_1 \mu_2, \nu_2} -
2 {\cal B}^{\mu_1 \mu_2, \nu}
\partial_{\mu_2} \partial^{\mu_3}
{\cal B}_{\mu_1 \mu_3, \nu}
\\ 
&-& 4 {\cal B}^{\mu_1 \mu_2, \nu} \partial_{\mu_1} \partial_{\nu} {\tilde M}_{\mu_2}
- \frac{2}{3}  {\tilde M}^\mu \, \Box \, {\tilde M}_\mu +
\frac{2}{3} {\tilde M}^\mu \, \partial_\mu \partial_\nu \,  {\tilde M}^\nu
\eea
 and is invariant under transformations
\bea \label{mst-1}
\delta {\cal B}_{\mu_1 \mu_2, \nu} &=& \partial_{(\mu_1} \Xi_{\mu_2) \nu}
 + \frac{2}{3}\partial_\nu
{ \Lambda}_{\mu_1 \mu_2} - \frac{1}{3} \partial_{(\mu_1} 
\Lambda_{\mu_2) \nu}  \\ 
\delta {\tilde M}_{\mu}& = & -\partial^\nu \Lambda_{\nu \mu} +
3\partial^\nu \Xi_{\nu \mu} + 3 \partial_\mu \sigma
\eea
with one symmetric $\Lambda_{\mu \nu}$
and one antisymmetric
$\Xi_{\mu \nu}$
parameters of gauge transformations.
  The new parameters  are related to the original ones 
  given in \p{loparameters},
  as
 \be 
 \tilde \Lambda_{\mu \nu} = \frac{1}{3}(\rho_{\mu \nu}+ \lambda_{(\mu, \nu)} ), \quad
 \Lambda_{\mu \nu}  =  \rho_{\mu \nu}- \frac{1}{2} \lambda_{(\mu, \nu)}, \quad
 \Xi_{\mu \nu}
  = \frac{1}{2} (\lambda_{\mu, \nu}- \lambda_{\nu, \mu})
 \ee
In order to decompose the Lagrangian 
\p{tr-3-1}
into a sum of the Lagrangians  for the individual Fronsdal modes
 one introduces the fields $\Phi_{ \mu_1 \mu_2 \mu_3}$ and $\Phi_{ \mu}$ as  
\be \label{db3}
\tilde \phi_{ \mu_1 \mu_2 \mu_3} = \Phi_{ \mu_1 \mu_2 \mu_3} + \frac{1}{d}  \eta_{(\mu_1 \mu_2} \Phi_{\mu_3)},
\quad
\Phi_\mu = \tilde \phi^\nu{}_{\nu \mu} - 2 \tilde D_\mu
\ee
As it was in the case  of $n=1$, the precise from of this decomposition is obtained from the requirement that the Lagrangian splits into a sum of Lagrangians for individual Fronsdal modes. However, one can get
the same result by requiring that the fields
$\Phi_{ \mu_1 \mu_2 \mu_3}$ and $\Phi_{ \mu}$ have 
``proper" gauge transformation rules i.e., 
\be
\delta \Phi_{ \mu_1 \mu_2 \mu_3}= \partial_{(\mu_1 } {\tilde \Lambda}_{\mu_2, \mu_3)},
\quad
\delta {\Phi}_{ \mu}= \partial_{\mu } \tilde \Lambda
\ee
where the parameter  $ \tilde \Lambda_{\mu \nu}$ is traceless.

For the system describing mixed symmetry 
 fields one performs the following field redefinitions
\be
{\cal B}_{\mu_1 \mu_2, \nu} = T_{\mu_1 \mu_2, \nu}
+ \frac{1}{2d-6}(2 \eta_{\mu_1 \mu_2} T_{\nu} 
- \eta_{\nu (\mu_1 } T_{\mu_2)}),  \qquad
T_\nu = {\cal B}^{\mu}{}_{\mu, \nu} - \frac{2}{3}{\tilde M}_\nu
\ee
This transformation results in  decomposition of the Lagrangian
\p{bpp} into a sum 
of a standard Lagrangian for a massless vector field
$T_\mu$ 
\be
{\cal L}^{s=1} = \frac{3}{2d-6} (T^\mu \, \Box \, T_\mu - T^\mu \partial_\mu \partial_\nu T^\nu )
\ee
being invariant under 
\be
\delta T_{\mu} = \partial_\mu \xi, \quad 
\xi= \frac{2}{3} (\Lambda^\mu{}_\mu-3 \sigma)
\ee
and
of the Lagrangian for the mixed symmetry field
\cite{Brink:2000ag} 
\bea \nonumber \label{BMV-ms}
{\cal L}^{(2,1)} &=&
 T^{\mu_1 \mu_2, \nu} \, \Box \,  T_{\mu_1 \mu_2, \nu} - 
T^{\mu_1 \mu_2, \nu_1}
\partial_{\nu_1} \partial^{\nu_2}
T_{\mu_1 \mu_2, \nu_2} -
2 T^{\mu_1 \mu_2, \nu}
\partial_{\mu_2} \partial^{\mu_3}
T_{\mu_1 \mu_3, \nu}
\\ 
&+& 3
T^{\mu}{}_{\mu, \nu}
\partial_{\mu_1} \partial_{\mu_2}
T^{\mu_1 \mu_2, \nu}
-\frac{3}{2} T^{\mu}{}_{\mu, \nu} \Box T^{\rho}{}_{\rho,}{}^ {\nu}
+ \frac{3}{2} T^{\mu}{}_{\mu, \nu_1} \partial^{\nu_1}
\partial^{\nu_2}
T^{\rho}{}_{\rho, \nu_2}
\eea
invariant under the same transformations as
\p{mst-1}, where the symmetric parameter 
is replaced with 
\be
\Lambda_{\mu \nu} \rightarrow  \Lambda_{\mu \nu}+\frac{1}{6-2d}
\eta_{\mu \nu}
\xi
\ee
 In the fermionic sector 
the   Lagrangian
\p{L-F} 
\bea \label{lf-1-1} \nonumber
{\cal L}_{f.}& = & 
 i \, \bar \Psi^{\mu_1 \mu_2} \gamma^\nu \partial_\nu \Psi_{\mu_1 \mu_2}
+ 2i \, \bar \Psi^{\mu_1 \mu_2}  \partial_{\mu_1} \chi_{\mu_2}
-2i \, \bar \chi_{\mu_1}  \partial_{\mu_2} \Psi^{\mu_1 \mu_2}
- 2i \, \bar \Sigma \, \gamma^\nu \partial_\nu \Sigma \\ 
&-& 2i \, \bar \chi^\mu \gamma^\nu \partial_\nu \chi_\mu
+2i \, \bar \chi^\mu \partial_\mu \Sigma - 2i \, \bar \Sigma \, \partial_\mu
\chi^\mu
\eea
is invariant under gauge transformations
\be
\delta \Psi_{\mu_1 \mu_2}= \partial_{\mu_1} \xi_{\mu_2}+
\partial_{\mu_2} \xi_{\mu_1}, \quad
\delta \chi_\mu = - \gamma^\nu \partial_\nu \xi_{\mu}, \quad
\delta \Sigma = \partial_\mu \xi^{\mu}
\ee
This Lagrangian can be converted into a sum of  Lagrangians,
each describing an 
individual Fang-Fronsdal mode
by using the transformations
\bea 
\tilde \Psi_{\mu_1 \mu_2} &=&
\Psi_{\mu_1 \mu_2}- \frac{1}{d} (\gamma_{(\mu_1} \gamma^\mu 
\Psi_{\mu_2) \mu} + \gamma_{(\mu_1} \chi_{\mu_2)}) -
\frac{1}{d} \eta_{\mu_1 \mu_2} ( \Psi^\mu{}_\mu- 2 \Sigma)  \label{52}\\ 
\tilde \Psi_{\mu} &=& \gamma^\nu 
\Psi_{\mu \nu} + \chi_\mu - \frac{1}{d-2} \gamma_\mu
 ( \Psi^\nu{}_\nu- 2 \Sigma ) , \label{32}\\ 
\tilde \Psi &=& \Psi^\mu{}_\mu- 2 \Sigma \label{12}
\eea
Similarly to the case of bosonic fields, the form of these transformations can be 
obtained from the requirement that
the new fields transform ``properly" under gauge transformations.
Namely, the expression \p{12}
 can be obtained from the requirement that $\tilde \Psi$  is
 the gauge invariant spin $\frac{1}{2}$ field.
For the spin $\frac{5}{2}$ and $\frac{3}{2}$ fields one has
the gauge transformation rules
 \be
\delta \tilde \Psi_{\mu_1 \mu_2} = \partial_{(\mu_1} \tilde \xi_{\mu_2)}, \quad \delta \tilde \Psi_{\mu} = \partial_\mu \tilde \xi,
\quad
 \tilde \xi_\mu = \xi_\mu - \frac{1}{d}\gamma_\mu \gamma^\nu \xi_\nu, \quad \tilde \xi = \gamma^\mu \xi_\mu
\ee
being satisfied by the transformations \p{52} and \p{32}.
Obviously, these considerations can not determine
the numerical coefficients in front 
of $\Psi$ in the last
terms of the equations \p{52} and \p{32},
 since  $\Psi$
 is  gauge invariant. The only possibility
 to determine
 these coefficients 
 is to require
that the Lagrangian \p{lf-1-1} when written
in terms of the irreducible modes, splits into a direct sum for individual irreducible representations of the Poincar\'e group.
 Noticing that there are always the same combinations of the ``untilded" fields
in \p{52}--\p{12},
it is straightforward
to invert these transformations
 by starting from the bottom one to obtain
\bea
 \Psi_{\mu_1 \mu_2} &=&
\tilde \Psi_{\mu_1 \mu_2} + \frac{1}{d} \gamma_{(\mu_1} \tilde \Psi_{\mu_2)} + \frac{1}{d-2} \eta_{\mu_1 \mu_2}\tilde \Psi \\ \nonumber
\chi_\mu &=& -\gamma^\nu \tilde \Psi_{\nu \mu} + \frac{1}{d}
(\gamma_\mu \gamma^\nu \chi_\nu - 2\chi_\mu) \\ \nonumber
\Sigma &=& \frac{1}{2}\tilde \Psi^{\mu}{}_{\mu} + \frac{1}{d} \gamma^\mu \tilde \Psi_\mu + \frac{1}{d-2} \tilde \Psi
\eea
Inserting these expressions into the Lagrangian
\p{lf-1-1}, one obtains a direct sum of Fang-Fronsdal Lagrangians
 for spins $\frac{5}{2}$, $\frac{3}{2}$ and $\frac{1}{2}$
with overall factors of $1$, $\frac{1}{d}$
and $\frac{1}{2(d-2)}$.

Finally,   using the field redefinitions
given in the present Section,  
and supersymmetry transformations
\bea \label{MAIN1-21}
&&\delta \Psi_{\mu_1 \mu_2} 
=-  \gamma^{\rho \nu}\,  \epsilon \, \partial_\rho 
\phi_{ \mu_1 \mu_2, \nu } + \epsilon \, \partial_{(\mu_1} B_{\mu_2)},
 \\ \nonumber
&&\delta \Sigma = - \gamma^{\rho \nu}
\, \epsilon \, \partial_\rho  D_{\nu }
+ \epsilon \partial^{\mu} B_{\mu},
 \\ \nonumber
&&\delta \chi_{\mu}=  \gamma^\nu\, \epsilon\,
(\partial^{\mu_1} \phi_{ \mu \mu_1,  \nu }
-\partial_\mu D_\nu  )+  \gamma^\nu\, \epsilon \partial_\nu A_\mu 
\eea
\be \label{MAIN2-21} \nonumber
\delta  \phi_{ \mu_1, \mu_2,  \nu } = i \, \bar \Psi_{\mu_1 \mu_2}
\gamma_\nu\, \epsilon,  \quad
\delta  D_{ \nu }
= i \, \bar \Sigma\gamma_\nu\, \epsilon, \quad
 \delta  B_{\mu} = -i \, \bar \chi_{ \mu} \, \epsilon, \quad
 \delta  A_{\mu} =0.
\ee
it is straightforward to obtain supersymmetry transformations
at each step described above.

Let us make a few comments on $AdS_d$ deformation for the case of  $n=2$.
Recall, that according to the so-called 
BMV conjecture  \cite{Brink:2000ag},
 performing an $AdS_d$ deformation
of the Lagrangian \p{BMV-ms} one loses
the gauge invariance with respect to the symmetric parameter
$\Lambda_{\mu \nu}$ and as a result the number of on-shell degrees of freedom changes\footnote{See 
\cite{Boulanger:2008up}--\cite{Alkalaev:2011zv} for a proof of the BMV conjecture,  \cite{Basile:2016aen} for a discussion for the case of  the de Sitter background
and
\cite{Reshetnyak:2023oyj} for recent progress in the BRST approach.}. In order to avoid this issue, 
one introduces
an extra Stueckelberg-type
rank-two field, which
couples to the original mixed symmetry field in the Lagrangian.
This new field  transforms under the parameter  $\Lambda_{\mu \nu}$
and  also brings  `` its own" gauge symmetry with a parameter
$\xi_\mu$. Then one can prove, that after fixing all gauge symmetries,
the modified system contains the same degrees of freedom as the original Lagrangian on the flat background.
From the consideration above one can conclude, that
in order to repeat the procedure  described in 
Subsection \ref{adsdeformation-s}
for the case at hand,
one has to follow the option when one performs
the splitting of the original Lagrangian into irreducible representations prior to the $AdS_d$ deformation.
In this case
a coupling of the Stueckelberg-like field
to the symmetric part of  $\phi_{\mu_1 \mu_2, \nu}$ can be avoided,
and this keeps a possibility for the diagonalization procedure also on
$AdS_d$. A detailed study of this question, 
is beyond the scope of the present paper and  
 we leave it for a future  
work.

\section{Diagonalization for a general \texorpdfstring{$n$}{n}} \label{sgen}
\setcounter{equation}0

The procedure that was shown  in detail in the previous
two Sections can be generalized to the case of an arbitrary $n$.
Using the language of auxiliary oscillators, described in  Section \ref{secreducible},
one starts with the Lagrangian \p{NSNS}
and splits the field $|\phi \rangle$
into a sum 
\be \label{phn}
| \phi \rangle = | \tilde \phi \rangle + | {\cal B} \rangle 
\ee
where the vectors $| \tilde \phi \rangle$ and $| {\cal B} \rangle $
contain fields with symmetry of  $Y(n+1,0)$ and   $Y(n,1)$, respectively
\bea \label{osc-11}
&&| \tilde \phi \rangle =\frac{1}{(n+1)!} \tilde \phi_{\mu_1,..,\mu_n \mu_{n+1}}(x) \,\alpha^{\mu_1,+}...\alpha^{\mu_n,+} \psi^{\mu_{n+1},+}|0\rangle \\ \nonumber
&& |{\cal  B}\rangle  =
\frac{1}{n!} {\cal B}_{\mu_1,..,\mu_n,\nu}(x) \,\alpha^{\mu_1,+}...\alpha^{\mu_n,+} \psi^{\nu,+}|0\rangle, 
\eea
The form of the coupling of the field $| \phi \rangle$
to the fields $| D \rangle$ and $| M \rangle$ in the Lagrangian
\p{NSNS}
suggests the field redefinitions
\be \label{Dn}
| \tilde D \rangle = - | M \rangle + t | D \rangle, \quad 
| \tilde M \rangle = t^+ | M \rangle + 2 | D \rangle
\ee
where the vectors $| \tilde D \rangle$ and
$| \tilde M \rangle $ have a form similar to \p{osc-11}, namely
\bea \label{osc-12}
&&| \tilde D \rangle =\frac{1}{(n-1)!} \tilde D_{\mu_1,..,\mu_{n-1}}(x) \,\alpha^{\mu_1,+}...\alpha^{\mu_{n-1},+} |0\rangle \\ \nonumber
&& |{\tilde  M}\rangle  =
\frac{1}{(n-2)!} {\tilde M}_{\mu_1,..,\mu_{n-2},\nu}(x) \,\alpha^{\mu_1,+}...\alpha^{\mu_{n-2},+} \psi^{\nu,+}|0\rangle
\eea
After inverting these transformations
\be
|  D \rangle = \frac{1}{2} 
(1 - \frac{1}{N_{\alpha, \psi}+2}t^+ t)
| \tilde M \rangle + t^+ | \tilde D \rangle, \quad 
|  M \rangle = \frac{1}{N_{\alpha, \psi}+2} t | \tilde M \rangle - 2 | \tilde D \rangle
\ee
one can check that,
in terms of the newly introduced fields,
the Lagrangian has a form \p{sum},
where
\bea \label{nmbD} 
{\cal L}^\prime_{bos.}&=&- \langle \tilde \phi | l_0 - l^+ l - g^+ g| \tilde \phi \rangle 
- \langle \phi| l^+ g^+ | \tilde D\rangle 
- \langle \tilde D | g \, l| \tilde \phi \rangle
\\ \nonumber
&+&  \langle \tilde D |(2l_0 + l^+ l)(N_{\alpha, \psi}+2)| \tilde D\rangle  
\eea
which describes reducible representations of type
$Y(n+1,0)$ 
and the Lagrangian
\bea \label{nGenLagtildeD-1} 
{\cal L}_{bos.}^{\prime \prime}&=& -\langle {\cal B} | l_0 - l^+ l - g^+ g| {\cal B} \rangle -
\frac{1}{2}  \langle {\cal B} | l^+ \, l^+  |\tilde M \rangle 
-\frac{1}{2}\langle \tilde M |l \, l  | {\cal B}\rangle \\
\nonumber
&+& \frac{1}{4} \langle \tilde M |  2 \, l_0 (1 + \frac{1}{N_{\alpha, \psi}+2} t^+ t) 
-  {\cal T} (- l^+ l + g^+ g) {\cal T} \\ \nonumber
&+& \frac{2}{N_{\alpha, \psi}+2}({\cal T} g^+ l t + t^+ l^+ g {\cal T})
|\tilde M \rangle,  
\eea
where
\be
N_{\alpha, \psi} = \alpha^+ \cdot \alpha + \psi^+ \cdot \psi, \quad
{\cal T}= 1 - \frac{1}{N_{\alpha, \psi}+2} t^+ t
\ee
which describes reducible representations with mixed symmetry  of type $Y(n,1)$.
Therefore, the entire system is described
as a sum of the bosonic Lagrangians
\p{nmbD},  \p{nGenLagtildeD-1} 
and of the fermionic Lagrangian
\p{L-F}.

In order to simplify the subsequent analysis
let us notice, that for the case of totally symmetric
fields, 
one can equivalently reformulate the Lagrangian \p{nmbD}
in terms of the fields that depend only on the bosonic oscillators
 $\alpha^{ +}_\mu$ as
\be \label{nmbD-1} 
{\cal L}^\prime_{bos.}=- \langle \tilde \phi | l_0 - l^+ l | \tilde \phi \rangle 
- \langle \phi| l^+ l^+ | \tilde D\rangle 
- \langle \tilde D | l \, l| \tilde \phi \rangle
+  \langle \tilde D |2l_0 + l^+ l| \tilde D\rangle  
\ee
 which is invariant under gauge transformations
 $\delta | \tilde \phi \rangle  = l^+ |\tilde \lambda \rangle $
 and $\delta | \tilde D \rangle  = l |\tilde \lambda \rangle $.
Then the  third step for the case of totally symmetric
bosonic fields
goes as follows \cite{Fotopoulos:2009iw}.
One introduces  fields
\be \label{ws}
| {\cal W}_n^p \rangle = (M_{11})^p|\tilde \phi_n \rangle - p
(M_{11})^{p-1}|\tilde D_{n-2} \rangle, \quad \delta | {\cal W}_n^p\rangle 
= l^+ (M_{11})^{p-1} |\tilde \lambda \rangle
\ee
and takes an ansatz
for irreducible fields in a form of an expansion
\be \label{expa}
 |\Phi _n\rangle = \sum_{p=0}^{[\frac{n}{2}]} c_p \, (M^+_{11})^p \, | 
{\cal W}_n^p \rangle 
\ee
The expansion coefficients $c_q$ can be found
\be \label{cpb}
c_p= (-1)^p \frac{(d+2(s-p-3))!!}{(d+2(s-3))!!}
\ee
from the requirement of the zero double trace condition
on $ |\Phi _n\rangle $, which should be satisfied by a Fronsdal mode. 
Alternatively,
most of the coefficients $c_q$ (except the ones that multiply gauge-invariant combinations $| {\cal W}_n^0 \rangle$)
can also be found from the requirement that the
Fronsdal field has the ``proper" gauge transformations
$\delta |\Phi _n\rangle = l^+ | \Lambda_{n-1} \rangle$, with
the traceless gauge parameter.
In order to check that the Lagrangian
for reducible higher spin fields
splits into a sum of the Lagrangians, each describing an irreducible mode,
one has to invert the transformations \p{expa}
and insert the inverted expression into the Lagrangian \p{nmbD-1}.
The fact that this procedure diagonalizes
the Lagrangian  was explicitly checked
up to   spin four. However, as it follows 
from the discussion above, it is natural to expect that the diagonalization of Lagrangians works for a general value of  $n$.

For the fermionic fields one can proceed in an analogous way. Namely, after defining fields
\bea
| {\cal W}_n^{2p} \rangle & =& (M_{11})^{p} \, | \Psi_n \rangle
  -p
(M_{11})^{p-1} \,  | \Sigma_{n-2} \rangle
\\
| {\cal W}_n^{2p+1} \rangle & =& (M_{11})^{p} \, T \, | \Psi_n \rangle
+ (M_{11})^{p} \, | \chi_{n-1} \rangle -p \,
(M_{11})^{p-1} \, T \, | \Sigma_{n-2} \rangle
\eea
one looks for a solution for an irreducible mode $| \tilde \Psi_n \rangle$
in the form of the expansion
\be \label{expa-f}
| \tilde \Psi_n \rangle = \sum_{p=0}^{n} c_{p} \, (T^{+})^p
\, | 
{\cal W}_n^p \rangle 
\ee
These expressions generalize \p{df1} and 
\p{52}--\p{12}
  to an arbitrary value of $n$.
The coefficients $c_p$ can be determined 
\be \label{cpf}
 c_{2p-1}(n,d)= (-1)^p \frac{(d+2(s-p-2))!!}{(d+2(s-2))!!}=c_{2p}(n, d)
\ee
from the requirement,
that the field $| \tilde \Psi_n \rangle$ is triple gamma-traceless,
or alternatively from the requirement, that
the irreducible mode  transforms ``properly" under 
the gauge transformations with the traceless parameter (see \p{FFF-1}--\p{FFF-2}).

For mixed symmetry fields, the procedure is essentially the same,
with the exception that now one has two types of ${\cal W}$, due
to the two possible types symmetries contained in the field
 $|\tilde M \rangle$.
Bellow we shall illustrate this procedure for the case
when the field $| {\cal B} \rangle$ is of the type
$Y(3,1)$, a generalization to an arbitrary $n$ is obvious. 
Recall, that as we discussed above, 
 in order to find the coefficients $c_p$
 for the case of  
 totally symmetric bosonic and fermionic fields, one can use the conditions of zero double
trace and zero triple gamma--trace, respectively. The result is given by equations \p{cpb} and \p{cpf}
by taking $n=3$.
On the other hand, for the mixed symmetry field
${\cal B}_{\mu_1 \mu_2 \mu_3, \nu}$
the zero double trace condition is automatically satisfied
and therefore we proceed with the requirement
of the ``proper" gauge transformations for it. 

First, using the gauge transformation rules \p{NSNSGT},
one can write the gauge transformations
for the fields \p{phn} and \p{Dn}
as follows
\be \label{gtt1}
\delta | {\cal B} \rangle  = (l^+ - g^+ t )|\Xi \rangle,
\ee
\be
\delta | {\tilde M } \rangle  = (3l + (t^+ t-1) l )|\Xi \rangle
+ (3 g^+  + l^+ t^+) |\sigma \rangle
\ee
where the parameters of gauge transformations are
\be
|\Xi \rangle = \frac{1}{2} \Xi_{\mu_1 \mu_2, \nu}(x)
\alpha^{\mu_1, +} \alpha^{\mu_2, +} \psi^{\nu, +}|0 \rangle, \quad
|\sigma \rangle =  \sigma_{\mu} (x)
\alpha^{\mu, +} |0 \rangle
\ee
The parameter $\Xi_{\mu_1 \mu_2, \nu}$ is symmetrical with respect to its first two indices, but has no symmetry with respect to the third index. Notice also that the gauge transformation rule
\p{gtt1} is consistent with the symmetry of the field 
${\cal B}_{\mu_1 \mu_2 \mu_3, \nu}$ i.e., with
the condition $t  |{\cal  B}\rangle  =0$.
Next we define in analogy with
\p{ws} 
\be
|{\cal W}^{(s)}\rangle = - t^+ t M_{11}  |{\cal  B}\rangle
+ \frac{1}{4} t^+ t | {\tilde M } \rangle
\ee
\be
| {\cal W}^{(as)}\rangle =(-2  + t^+ t) 
 \, (M_{11} |{\cal  B}\rangle
- \frac{1}{2} | {\tilde M } \rangle), \quad
t | {\cal W}^{(as)}\rangle=0
\ee
which contain symmetrical and antisymmetrical tensors of the second rank, respectively.
These fields are defined in such a way, that similarly
to \p{ws}
 their gauge transformations  do not contain divergences of the parameters
 \be
 \delta |{\cal W}^{(s)}\rangle =
 (l^+ t^+ M_{12} + g^+ M_{12}) |\Xi \rangle +
(l^+ t^+ + g^+) |\sigma \rangle
 \ee
 \be
 \delta |{\cal W}^{(as)}\rangle =
 (-l^+ (2 M_{11} + t^+ M_{12})
 + g^+(2t M_{11} + M_{12})
 ) |\Xi \rangle +
(-l^+ t^+ + g^+) |\sigma \rangle
 \ee
Finally, we are looking for an irreducible
field ${\cal T}_{\mu_1 \mu_2 \mu_3, \nu}$
in the form 
\be
| {\cal T} \rangle =
| {\cal B} \rangle +
\frac{1}{2(d-2)}(2 M_{11}^+ -   M_{12}^+t)|{\cal W}^{(s)}\rangle
+  \frac{1}{d}M_{11}^+|{\cal W}^{(as)}\rangle
\ee
where the expansion coefficients 
can be found after recasting the gauge transformations for the fields
$| {\cal T} \rangle$ in the form \p{gtt1}, where
the parameter of gauge transformations is now traceless \cite{Burdik:2000kj}.

\section{Conclusion}

There are several directions in which our results may be applied. In particular, the diagonalization procedure seems to be an essential tool in the construction of 
supersymmetric Lagrangians which contain reducible higher spin fields on an $AdS_d$ background, a problem that we hope to address in a separate publication.  
It would also be interesting to apply the diagonalization procedure
developed in the present paper to
supersymmetric models which contain cubic and higher-order interactions
 \cite{Buchbinder:2017nuc}--\cite{Buchbinder:2022kzl}
either on flat or on $AdS_d$ backgrounds
similar to how it has been done for the case of current-current interactions \cite{Fotopoulos:2009iw}--\cite{Fotopoulos:2010nj}.

Another interesting possibility would be to study applications
of the diagonalization procedure to quantum supersymmetric higher spin theories, (see  
\cite{Steinacker:2022jjv}--\cite{Steinacker:2024huv} for recent progress in supersymmetric quantum higher spin theories).

Finally, let us mention
that the diagonalization procedure has not been studied
for the case of reducible  massive higher spin fields
\cite{Burdik:2000kj}, even in the absence of supersymmetry.
A possible progress in this direction can prove to be useful for the field-theoretic description of interacting spinning back holes,  constructed in terms of massive and massless higher spin triplets \cite{Skvortsov:2023jbn}.

\vskip 0.5cm

\noindent {\bf Acknowledgments.} We would like to thank
Sergei Kuzenko and Yasha Neiman for discussions.   The work was supported by the Quantum Gravity Unit of the Okinawa Institute of Science and Technology Graduate University (OIST). 

\renewcommand{\thesection}{A}

\renewcommand{\theequation}{A.\arabic{equation}}

\setcounter{equation}0
\appendix
\numberwithin{equation}{section}

\section{Lagrangians and the algebra of operators on AdS  space}\label{Appendix A}
Although in this paper we were primarily concerned with a flat
background, here we give the description of massless bosonic and fermionic higher spin fields for a $d$-dimensional anti-de Sitter space, originally given for $d=4$ in 
\cite{Fronsdal:1978vb}--\cite{Fang:1979hq}. 
The field equations for general mixed symmetry fields
on $AdS_d$ are given in \cite{Metsaev:1997nj}.
The corresponding equations for the Minkowski space-time can be simply recovered by taking the limit $L \rightarrow \infty$,
where $L$ is the radius of the $AdS_d$ space.

We use the language of an auxiliary Fock space, given in Section
\ref{secreducible}. The indexes $\mu, \nu,...$ are Einstein indices and
$a,b,...$ are tangent space indices. 
We use the notation $A \cdot B \equiv A^\mu B_\mu$. 
We define symmetrization without a factor in the denominator, for example
 $A_{(\mu \nu)}= A_{\mu \nu} + A_{\nu \mu}$.

The metric is mostly plus.
The $AdS_d$ covariant derivatives are realized as
\be \label{c-1}
p_\mu = -i (\partial_\mu + \omega_\mu{}^{ab} \alpha_a^+ \alpha_b), 
\qquad
[\alpha_a, \alpha_b^+] = \eta_{ab}
 \ee
\be \label{c-2}
[p_\mu, p_\nu] = \frac{1}{L^2} (\alpha_\mu^+ \alpha_\nu
- \alpha_\nu^+ \alpha_\mu) + \frac{1}{2L^2}\gamma_{\mu \nu}
\ee
where the last term in \p{c-2} appears when acting on spinors
and $\gamma^{\mu \nu}= \frac{1}{2} (\gamma^\mu \gamma^\nu - \gamma^\nu \gamma^\mu)$.
Further, we introduce operators
\be \label{ope-1}
l_0 = g^{\mu \nu} (p_\mu p_\nu + i \Gamma^\lambda_{\mu \nu} p_\lambda), \quad T_0 = \gamma \cdot p
\ee
\be
l= \alpha \cdot p, \quad  l^+= \alpha^{+} \cdot p,
\quad T = \alpha \cdot \gamma, \quad T^+= \alpha^{+} \cdot \gamma
\ee
and 
\be \label{ope-3}
N_\alpha = \alpha^{ +} \cdot \alpha, \quad M_{11}^+ = \frac{1}{2}
\alpha^{ +} \cdot \alpha^+, \quad 
M_{11} = \frac{1}{2}
\alpha \cdot \alpha,
\ee
The action of these operators on a state $|\Phi \rangle $
is translated to the action on a symmetric (spinor) tensor field
\be \label{st-1}
|\Phi \rangle = \frac{1}{s!} \Phi_{\mu_1,...,\mu_s}(x) \alpha^{\mu_1,+}...\alpha^{\mu_s,+}|0 \rangle
\ee
as follows
\be \label{ac-1}
l_0 |\Phi \rangle \rightarrow - \Box \, \Phi_{\mu_1,...,\mu_s}(x), \quad
T_0 |\Phi \rangle \rightarrow  - \gamma^\mu \nabla_\mu \Phi_{\mu_1,...,\mu_s}(x)
\ee
\be
l |\Phi \rangle \rightarrow  - i \nabla^{\mu_1} \Phi_{\mu_1,...,\mu_s}(x), \quad
l^+ |\Phi \rangle \rightarrow  - i \nabla_{(\mu} \Phi_{\mu_1,...,\mu_s)}(x)
\ee
\be
T |\Phi \rangle \rightarrow   \gamma^{\mu_1} \Phi_{\mu_1,...,\mu_s}(x), \quad
T^+ |\Phi \rangle \rightarrow   \gamma_{(\mu} \Phi_{\mu_1,...,\mu_s)}(x)
\ee
\be \label{ac-5}
M^+_{11} |\Phi \rangle \rightarrow   g_{(\mu \nu} \Phi_{\mu_1,...,\mu_s)}(x), \quad
M_{11} |\Phi \rangle \rightarrow  \frac{1}{2} \Phi^\mu{}_{\mu,...,\mu_s}(x)
\ee
\be \label{ac-6}
N_\alpha |\Phi \rangle \rightarrow   s \Phi_{\mu_1,...,\mu_s}(x)
\ee
where $\nabla_\mu$ and $\Box$ are
 $AdS_d$ covariant derivative  and a 
D`Alembertian, respectively. 
 Using the commutation relations \p{c-1}--\p{c-2}
 it is straightforward to compute
 \bea \label{comu1}
 [l, l^+]  &=&  l_0 - 
 \frac{1}{L^2} (4 M^+_{11} M_{11}  - N^2_\alpha + 2N_\alpha -dN_\alpha) 
-\frac{1}{2L^2}\alpha^{\mu, +} \alpha^\nu  \gamma_{\mu \nu} \\ \nonumber
&\equiv& \tilde l_0  -\frac{1}{2L^2}\alpha^{\mu, +} \alpha^\nu \gamma_{\mu \nu}
 \eea
 \be
 \{ T_0, T_0 \} = 2 l_0 + \frac{2}{L^2} (T^+ T - N_\alpha)
 + \frac{1}{2L^2}
 (-d^2 +d  )
 \ee
 \be \label{comu3}
 [T_0, l] = \frac{1}{L^2} (2 T^+ M + \frac{1}{2}(1-2N_\alpha -d)T)
 \ee
 as well as
 \be \label{comu4}
 [l, M^+] = l^+, \quad
 [N_\alpha, l] = l, \quad [N_\alpha, M_{11} ] = 2M_{11},
 \quad [M_{11}, M^+_{11} ] = N_\alpha + \frac{d}{2}
 \ee
 \be \label{comu5}
 \{ T, T \} = 4M_{11}, \quad [T, l^+] = T_0, \quad   \{ T, T_0 \}= 2l, \quad
 [T, M^+_{11} ] = T^+
 \ee
 \be \label{comu6}
 \{ T, T^+ \} = 2N_\alpha +d , \quad [N_\alpha, T] = T
 \ee
The Lagrangian for an irreducible massless higher spin field $|\Phi \rangle $ in a $d$ dimensional anti-de Sitter space reads
\cite{Buchbinder:2001bs}
\bea \label{Fronsdal}
{\cal L}_{bos.} &=&- \langle \Phi |
\tilde l_0 - l^+ l - 2 M^+_{11}  \tilde l_0 M_{11}  + M^+_{11}  l \, l
+ l^+ l^+ M_{11}  - M^+_{11}  l^+ l M_{11}  
\\ \nonumber
&+&\frac{1}{L^2}  ( 6 - 4   N_\alpha  -2d  + 
 M^+_{11}   ( 10- 4N_\alpha - 2 d  )  M_{11}   ) | \Phi \rangle.
\eea
Using \p{comu1} (without the last term in the r.h.s.) and \p{comu4} one can check,
that  the Lagrangian is invariant under gauge transformations
\be
\delta | \Phi \rangle = l^+ |\Lambda \rangle
\ee
The field $| \Phi \rangle$ and the parameter of gauge transformations
$| \lambda \rangle$ obey off-shell constraints
\be
(M_{11})^2  |\Phi \rangle =0, \quad M_{11}  |\Lambda \rangle =0 
\ee
The Lagrangian for a massless half-integer higher spin field
on $AdS_d$ is \cite{Francia:2007qt}--\cite{Buchbinder:2007vq}
\bea \label{Fang-Fronsdal} \nonumber
{\cal L}_{f.} &=&- \langle \Psi |
 T_0 - l^+ T - 
 T^+ l + T^+ T_0 T + l^+ T^+ M_{11}  + M^+_{11}  Tl - M^+_{11} T_0 M_{11}  
\\ 
&-&\frac{i}{2L}  ( 2N_\alpha-4 + d + T^+ (2-2N_\alpha-d)T
- \\ \nonumber
&-&M^+_{11}  (2N_\alpha +d)M_{11} ) 
| \Psi \rangle
\eea
Using \p{comu1} --- \p{comu6} one can check that it is
invariant under gauge transformations
\be \label{FFF-1}
\delta | \Psi \rangle = (l^+ - \frac{i}{2L} T^+) | \xi \rangle
\ee
with off-shell constraints on the field 
$| \Psi \rangle$ and on the
parameters of gauge transformations $| \xi \rangle$ being
\be \label{FFF-2}
T^3| \Psi \rangle =0, \quad T| \xi \rangle =0
\ee

Finally, when considering mixed symmetry fields on a flat background,
in addition to the previously introduced operators
(with $p_{\mu} = -i \partial_\mu$ and $L \rightarrow \infty$)
we also use
\be
g= \psi \cdot p, \quad g^+= \psi^{+} \cdot p
\ee
\be
 N_{\alpha, \psi} = \alpha^+ \cdot \alpha + \psi^+ \cdot \psi, \quad
\quad M_{12}^+ = 
\alpha^{ +} \cdot \psi^+, \quad 
M_{12} = 
\alpha \cdot \psi,
\ee
 \be \label{deft}
 t^+ = \psi^+ \cdot \alpha, \quad t =   \alpha^+ \cdot \psi, \quad
\ee 
The states are now either of  the form \p{st-1}
or
\be \label{st-2}
|\tilde \Phi \rangle = \frac{1}{s!} \tilde \Phi_{\mu_1,...,\mu_s, \nu}(x) \alpha^{\mu_1,+}...\alpha^{\mu_s,+}\psi^{\nu,+}|0 \rangle
\ee
The action of the operators \p{ope-1}--\p{ope-3}
on states \p{st-2}
are the same as
  \p{ac-1}--\p{ac-6}. We also have
\bea
&&g |\tilde \Phi \rangle \rightarrow  - i \partial^{\nu} \tilde \Phi_{\mu_1,...,\mu_s,\nu}(x), \qquad \,\,\,\,\,
g^+ |\Phi \rangle \rightarrow  - i \partial_{\nu} \Phi_{\mu_1,...,\mu_s}(x)
\\
&&t |\tilde \Phi \rangle \rightarrow   \tilde \Phi_{(\mu_1,...,\mu_s,\nu)}(x), \qquad \,\,\,\,\,\,\,\,\,\,
\,\,\,\,\,
t^+ |\Phi \rangle \rightarrow   \Phi_{\mu_1,...,\nu}(x)
\\
 \label{ac-5-1}
&&M_{12} |\tilde \Phi \rangle \rightarrow   \tilde \Phi^\nu{}_{\mu_2,...,\mu_s, \nu}(x), \qquad \,\,\,\,\,\,
M_{12}^+ |\Phi \rangle \rightarrow  
g_{\nu (\mu_1} 
\Phi_{\mu_2,...,\mu_s)}(x)
\\
 \label{ac-6-1}
&&N_{\alpha, \psi} |\tilde \Phi \rangle \rightarrow   (s+1) \tilde \Phi_{\mu_1,...,\mu_s, \nu}(x), \qquad
N_{\alpha, \psi} |\Phi \rangle \rightarrow   s \Phi_{\mu_1,...,\mu_s}(x), 
\eea

\end{document}